# Intrusion detection systems using classical machine learning techniques versus integrated unsupervised feature learning and deep neural network

Shisrut Rawat[1], Aishwarya Srinivasan[2], and Vinayakumar R[3]

[1] Self-employed Data Scientist, New Delhi, India. shisrut.rawat@gmail.com
[2] Data Scientist, IBM, New York, USA. aishgrt@gmail.com,
[3] Researcher, Center for Computational Engineering and Networking (CEN), Amrita School of Engineering, Amrita Vishwa Vidyapeetham, Coimbatore, Tamil Nadu, India. vinayakumarr77@gmail.com

**Abstract.** Security analysts and administrators face a lot of challenges to detect and prevent network intrusions in their organizations, and to prevent network breaches, detecting the breach on time is crucial. Challenges arise while detecting unforeseen attacks. This work includes a performance comparison of classical machine learning approaches that require vast feature engineering, versus integrated unsupervised feature learning and deep neural networks on the NSL-KDD dataset. Various trials of experiments were run to identify suitable hyper-parameters and network configurations of machine learning models. The DNN using 15 features extracted using Principal Component analysis was the most effective modeling method. The further analysis using the Software Defined Networking features also presented a good accuracy using Deep Neural network.

**Keywords:** Intrusion Detection · Machine Learning · Deep Learning · Unsupervised learning · Dimensionality Reduction · Software Defined Networking.

## 1  Introduction

With the era of digitization and the Internet of Everything where all devices are paired into a signal network of communication, network attacks and endpoint attacks have splurged to a vast extent. Cybersecurity involves techniques and technologies to protect the device's software network, and data from unauthorized and unauthenticated access, malware attacks and network attacks [8]. Multiple systems have been designed around each of these spaces targeting specific detection and prevention methodology. This paper revolves around network intrusion attacks, classical and rule-based methods, recent advancements using machine learning and a proposal of a two-level model integrating unsupervised and deep neural networks. The effectiveness of the network intrusion detection the system comes into play when apart from identifying the known attacks; the system can detect inherited and new attacks. The thumb of rule-based NIDS are broadly classified into misuse-based or signature-based (SNIDS), anomaly-based (ANIDS) and ensemble methods. In signature-based NIDS the attack signatures are hardcoded and matching of these patterns is performed for incoming traffic to catch any abnormal traffic in the network. In, anomaly-based NIDS abnormal traffic is flagged; it is well designed for the recognition of new patterns of abnormal traffic. It is one of the most efficient to detect zero-day attacks which are not well supported using SNIDS. However, the performance of ANIDS in terms of false-positive rate is very high [8]. These two systems can be well integrated leveraging the strength points of SNIDS and ANIDS.

Machine learning capabilities have been seen in many domains and data is the powering factor behind these algorithms deep learning method, specialized machine learning approach is capable of internally doing the feature engineering with activation being optimized in the nodes of its multiple layers deep learning hence is an optimal method to compare the distribution of input data and hence proves to be a better method to detect zero-day attacks. Use of deep learning algorithms to various cybersecurity application such as malware analysis, intrusion detection, and botnet detection has improved the results significantly [1]. In this paper, ML and DL models are trained on the NSL-KDD data set and various performance matrix are compared. Additionally, a NIDS is designed and tested exclusively based on software-defined networking.

The paper starts with the related work for intrusion detection using network service access using machine learning techniques and the advancements in the methods in Section 2. The paper follows by details about the dataset used for the analysis in Section 3. The methodology section describes the details of the models built for intrusion detection in Section 3. The study presents a comparative analysis of multiple machine learning models versus deep learning models in Section 4. Conclusion and Future works are placed in Section 5.

## 2   Related Work

A self-taught learning-based NIDS is proposed in [2], where a sparse autoencoder and softmax regression is used. The proposed model is trained on the NSL-KDD dataset and it achieves an accuracy around 79.10% for 5-class classification which is very close to the performance of other state-of-the-art models. Apart from this, 23-class and 2-class classification are also achieving good performance. In [3], the performance of RNN based NIDS is studied. The model is trained on the NSL-KDD dataset, binary and multi-class classification are performed. The performance of RNN based IDS is far superior in both classification when compared to other traditional approaches and the author claims that RNN based IDS has a strong modeling capability for IDS. Unlike the above works, [4] proposes IDS for the SDN environment. A DNN based model is trained on only six basic features taken from the NSL-KDD dataset with different learning rates and it achieves a maximum accuracy of 75.75%. In [5], a new stacked nonsymmetric deep autoencoder (NDAE) based NIDS is proposed. The model has trained on both KDD Cup 99 and NSLKDD benchmark datasets and its performance is compared with DBN based model. It can be observed from the experimental analysis that the NDAE based approach improves the accuracy of up to 5% with 98.8% training time reduction when compared to DBN based approach. In [6], the authors have claimed that modeling network traffic data as a time series improves the performance of IDS. They substantiate the claim by training LSTM models with the KDD Cup dataset with a full and minimal feature set for 1000 epochs and have obtained a maximum accuracy of 93.82%. In [7], the effectiveness of CNN and CNN-RNN based models are studied. Models such as CNN, CNN-LSTM, CNN-GRU, and CNN-RNN are trained on the KDD Cup dataset and it can be observed that CNN based model outperforms hybrid CNN-RNN models. Unlike previously mentioned works, [8] analyses several ML-based approaches for intrusion detection for identifying various issues. Issues related to the detection of low-frequency attacks are discussed with a possible solution to improve the performance further. In [9], a highly scalable deep learning framework is proposed for intrusion detection at both network and host level. Various ML and DNN models are trained on datasets such as KDD Cup, NSLKDD, WSN-DS, UNSW-NB15, CICIDS 2017, ADFA-LD and ADFA-WD and their performance is compared.

## 3 Methodology

### 3.1 Description of Dataset

The network security datasets are available in two ways, First, from packet monitoring software such as Wireshark, Tcpdump, WinDump etc but these data will not be labelled and a lot of time will go into labeling hence may not be suitable for modelling purposes but can serve the purpose of an out time validation data set in that ensures the robustness of the ML/DL model. Second way is the use of open-source network security datasets available for free download, it saves data acquisition time and increases efficiency of research because they require very less cleaning and are present in a condition suitable for a modeler, For example DARPA Intrusion detection dataset, KDD Cup 99 dataset, ADFA dataset, NSL KDD dataset [8]. For our research used the NSL KDD dataset [10], it is a better version of the KDD Cup 99 dataset. One of the major drawbacks with the KDD Cup 99 dataset is a large number of duplicate observations in test and train, the NSL KDD dataset overcomes these limitations hence, it suits our purpose of building robust predictive models.

For each observation in the NSL KDD dataset, there are 41 features,3 are nominal, 4 are binary and the remaining 34 are continuous variables. It has 23 traffic classes in the training dataset and 30 in the test dataset. These attacks can be clustered into four main categories DOS, probing, U2R and R2L. The features are classified into 3 broad types 1) basic features, 2) content-based features and 3) traffic-based features. The attack information of the NLS-KDD dataset is listed in Tables 1 and 2.

**Table 1.** Dataset network intrusion details.

| Traffic | Train | Test |
|---------|-------|------|
| Normal | 67,343 | 9,711 |
| Dos | 45,927 | 7,458 |
| U2R | 52 | 67 |
| R2L | 995 | 2,887 |
| Probe | 11,656 | 2,421 |

**Table 2.** Subcategories of intrusions under each broader class intrusion (The highlighted attacks are only present in the test dataset).

| Category | Attacks |
|----------|---------|
| DoS | back, land, neptune, pod, smurf, teardrop, mailbomb,processtable, udpstorm,apache2,worm |
| R2L | fpt-erite,guess-passwd, imap, multihop, phf, spy, warezmaster, xlock,xsnoop,snmpguess,snmpgetattack,httptunnel,sendmail,named |
| U2R | buffer-overflow, loadmodule, perl, rootkit, sqlattack, xterm, ps |
| Probe | ipsweep, nmap, portsweep, satan, mscan, saint |

## 4   Model architecture

The proposal includes an unsupervised feature selection combined with the deep neural network, shown in Figure 1 and a deep neural network without unsupervised feature selection is shown in Figure 2. Following the hyperparameter selection study, the Deep Neural Network of 5-layers was created. The deep neural network is an advanced model of classical feed-forward network (FNN). As the name indicates the DNN contains many hidden layers along with the input and output layer. When the number of layer increases in FFN causes the vanishing and exploding gradient issue. To handle the vanishing and exploding gradient issue, the ReLU non-linear activation was introduced. ReLU helps to protect weights from vanishing by the gradient error. Compared to other non-linear functions, ReLU is more robust to the first-order derivative function since it does not zero for high positive and negative values of the domain. The proposed DNN architecture contains an input layer, 5 hidden layers, and an output layer. The output layer of DNN contains sigmoid activation function with a unit, which results in either 0 or 1. The value 0 indicates normal and 1 indicates an attack. The DNN model uses binary cross-entropy as loss function that can be defined as follows

$$loss(p, e) = -\frac{1}{N} \sum_{i=1} [e_i log(p_i) + (1 - e_i)log(1 - p_i)] \quad (1)$$

**p** = predicted labels vector, **e** = truth/expected label vector.

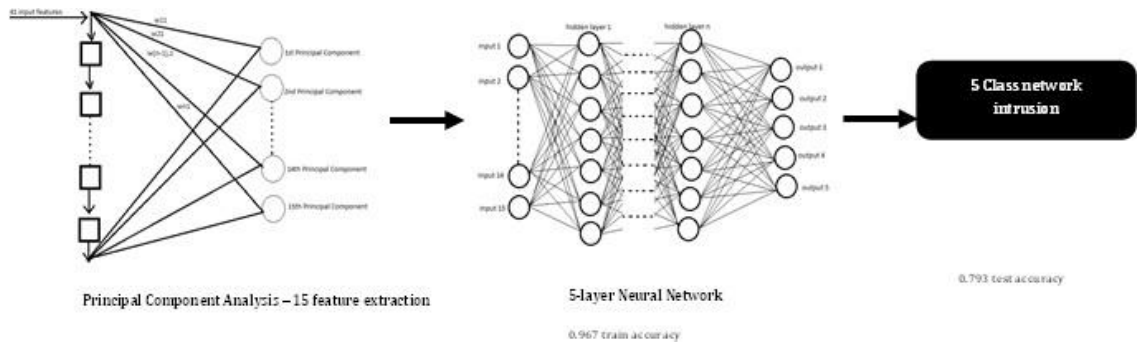

**Fig. 1.** Model architecture with 41 features.

## 5   Evaluation & Results

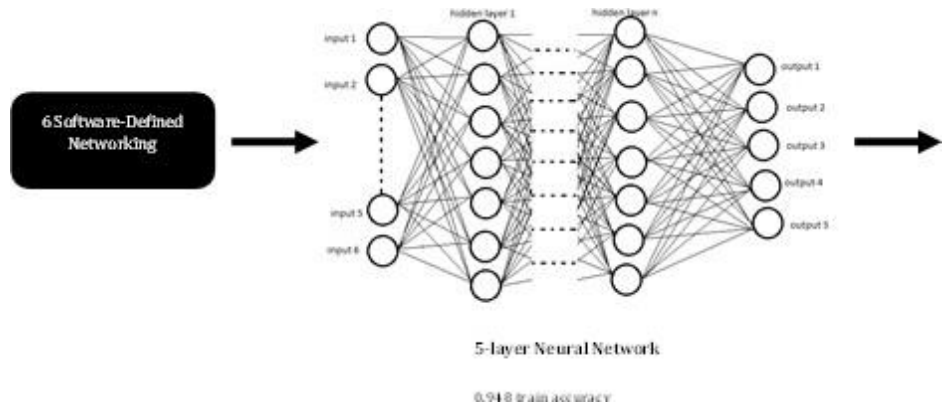

**Fig. 2.** Model architecture with 6 Software Defined Networking features.

Deep neural networks (DNNs) were trained using GPU enabled Tensor- Flow[5] as backend with Keras[6] framework. To compare the performance of various models using the NSL-KDD dataset, the following different scenarios were taken into consideration.

1. Classification of the network connection records as normal or attack considering all features present in the NSL-KDD dataset.
2. Classification of the network connection records as normal or attack considering minimal feature present in the NSL-KDD dataset.

The network connection records in the dataset are either Normal or Attack in the case of binary classification.

1. TP - connections that were accurately classified as the Normal class. (True Positive)
2. TN - connections that were accurately classified as the Attack class. (True Negative)
3. FP - Normal connection inaccurately classified as the Attack connection. (False Positive)
4. FN - Attack connection inaccurately classified as the Normal connection. (False Negative)

---

[5] https://www.tensorflow.org/
[6] https://keras.io/

Accuracy: It the ratio of the accurately classified network connections to the entire test dataset. Larger the accuracy better the classification model, the range of accuracy score is between 0 and 1. Accuracy score is defined as follows

$$Accuracy = \frac{TP + TN}{TP + TN + FP + FN} \qquad (2)$$

**Table 3.** Model performance with all features.

| Algorithm | Train Accuracy | Validation Accuracy | Test Accuracy |
|---|---|---|---|
| Decision Tree | 1.0 | 0.9978 | 0.778 |
| Extra Tree | 1.0 | 0.9973 | 0.767 |
| Ensemble Extra Tree | 1.0 | 0.999 | 0.769 |
| Light GBM | 0.996 | 0.989 | 0.776 |
| Deep Neural Network | 0.949 | 0.972 | 0.772 |
| PCA + Deep Neural Network | 0.967 | 0.982 | 0.793 |

**Table 4.** Model performance with 6 SDN features.

| Algorithm | Train Accuracy | Validation Accuracy | Test Accuracy |
|---|---|---|---|
| Decision Tree | 0.978 | 0.975 | 0.712 |
| Extra Tree | 0.978 | 0.973 | 0.744 |
| Ensemble Extra Tree | 0.978 | 0.974 | 0.736 |
| Light GBM | 0.976 | 0.966 | 0.742 |
| Deep Neural Network | 0.948 | 0.955 | 0.759 |

The models built for the study include training Decision Tree, Extra Tree, Ensemble Extra Tree, and Light GBM and DNN. In addition to the analysis, instead of using all features as the input to the DNN, Principal Component Analysis was applied on the 41 features to extract 15 reduced features and then fed into DNN. The hyperparameters were tuned for all the aforementioned models, whose details are not explicitly mentioned in the paper. All the models were run on train data of NSL-KDD with stratified cross-validation and later tested on the test data of NSL-KDD. As mentioned in the model architecture section, the models were trained and tested on 41 features and 6 features separately. According to multiple types of research by Tang [4], the intrusion dataset consists of six features that depict the Software Defined Networking features, namely duration, protocol type, source byte, destination byte, same host connection, and same service connection. To observe the relative performance of the predictive model over using all intrusion features versus SDN features, the models were built using just these 6 features. The results from the models on train, validation and test sets are presented in Tables 3 and 4 for the NSL-KDD dataset with 41 features and NSL-KDD dataset with minimal feature sets. The classical models performed better than the DNN on NSL-KDD dataset with 41 features. However, The DNN model performances better than the classical modes with minimal feature sets. Also, the performance attained by all the models with minimal feature sets is closer to 41 feature sets of the NSL-KDD dataset. This infers that all 41 features are not significant and most importantly the DNN model performed better on the reduced dataset.

## 6  Conclusion and Future work

In this paper, a deep learning algorithm for intrusion detection in networks was implemented and evaluated. As seen in the test dataset, there are multiple new intrusions were seen within each broader category. When the model was trained and evaluated on the train-validation split, the model performance was quite high, compared to test set accuracy, where new intrusions are seen. Compared to all other classifiers, the deep neural network presents a much better model fitting and better accuracy on the test set with a 0.793 accuracy. The other models seem to overfit the training data while performing less effectively on recognizing the intrusion patterns in the test data. Another implementation focuses on the Software Defined Networking variables for model training and evaluation. With just the 6 features out of the 41 features, the deep learning model gives an accuracy of 0.759 on the test set with unseen intrusions. In the future, we plan to implement a continuous real-time model training to have better performance rather than model training on static data.